# GerAPlanO – A NEW BUILDING DESIGN TOOL: DESIGN GENERATION, THERMAL ASSESSMENT AND PERFORMANCE OPTIMIZATION

**Eugénio Rodrigues**[1,2]\*, **Ana Rita Amaral**[1],
**Adélio Rodrigues Gaspar**[1], **Álvaro Gomes**[2,3],
**Manuel Carlos Gameiro da Silva**[1] **and Carlos Henggeler Antunes**[2,3]

1: ADAI, LAETA & Department of Mechanical Engineering,
Faculty of Sciences and Technology,
University of Coimbra
Rua Luís Reis Santos, Pólo II, 3030-788 Coimbra, Portugal
\* e-mail: eugenio.rodrigues@gmail.com, web: http://www.uc.pt/en/efs/research/geraplano

2: INESC Coimbra – Institute for Systems Engineering and Computers at Coimbra
Rua Antero de Quental 199, 3000-033 Coimbra, Portugal

3: Department of Electrical and Computer Engineering,
Faculty of Sciences and Technology,
University of Coimbra
Pólo II, 3030-290 Coimbra, Portugal

**Keywords:** software package, architectural design, building performance, thermal comfort, optimization, energy efficiency

**Abstract** *Building practitioners (architects, engineers, energy managers) are showing a growing interest in the design of more energy efficient and livable buildings. The best way to predict how a building will behave regarding energy consumption and thermal comfort is to use a dynamic simulation tool. However, the use of this kind of tools is difficult on a daily basis practice due to the heuristic and exploratory nature of the architectural design process. To deal with this difficulty, the University of Coimbra and three companies have been working on the development of a prototype design aiding tool, specifically devoted to the space planning phase of building design, under the project GerAPlanO (Automatic Generation of Architecture Floor plans with Energy Optimization). This project aims to combine the capabilities of design generation techniques, thermal assessment programs, and design optimization methods to provide assistance to decision makers. This paper presents the overall concept, as well as the current status of development of this tool.*

1. INTRODUCTION

The efficient use of energy in the built environment is a very important issue for a more sustainable future. The building stock accounts for 31% of the global energy consumption, with space heating and cooling representing one-third of that value (up to 60% in cold climates) [1]. This end-use consumption can be reduced if buildings have more adequate designs. Therefore, building performance simulation software should be included in the design process at the outset, since energy efficient solutions are easier to incorporate at the early phase of building design. The use of performance-based design tools represents an important methodological shift in the architectural design paradigm [2]. However, it is difficult predicting and incorporating energy efficiency concerns in the building design process, as it requires a long time to build accurate simulation models and understand which design options are more adequate in each design stage [3]. To deal with this problem, University of Coimbra's researchers have been working in the development of new building design tools aimed at a specific architectural design tasks in the framework of the GerAPlanO Project (Automatic Generation of Architecture Floor plans with Energy Optimization) [4]. The foremost purpose of this software package is to assist practitioners in the earliest design stage of architectural project, the space planning. In this phase, architects – the main professional target of these tools and henceforth considered as the users – seek to synthetize their options and requirements with the client's desires and aspirations into floor plans. The architectural design method still mostly consists in manual trial-and-error processes, based on past experience and rule-of-thumb actions; often the achieved solutions do not have the opportunity of being validated or estimated. In this sense, the aim of this approach is not to replace the architects' role in design and creativity, but to help in their decision-making process. This is done by providing useful energy performance data or thermal improvements based on space organization, which otherwise would be neglected or discarded due to the lack of time to introduce changes in late design phases or lack of knowledge on buildings energy simulation.

GerAPlanO tools generate alternative solutions for floor plans, which are then assessed and optimized. A software prototype resulting from this research [5], which has been developed under the *Energy for Sustainability Initiative* of the University of Coimbra, is dedicated to space planning, which consists in determining the best configuration of indoor spaces. This is a co-promotion project between the University of Coimbra and three companies – VisioArq, an architectural office; CiberBit, a software development company; and WSBP, a building technologies company. The aim is to develop a tool capable of assisting practitioners in generating, evaluating and optimizing alternative floor plan design solutions. The project began in September 2013 and will be finalized in June 2015. At this stage the prototype of a commercial software tool is being concluded, and the study of real case scenarios in architectural practice context started in February 2015.

## 2. GERAPLANO PROJECT

### 2.1. Design generation

The GerAPlanO tools generate alternative designs by organizing the indoor spaces according to the user's specifications for each storey. The spaces are arranged using a hybrid evolutionary strategy technique that takes into account a set of objectives and constraints [6–8]. Fifteen performance indicators are grouped into three main categories: Floor plan - satisfying construction and gross areas limits, minimizing circulation areas, maximizing construction and gross areas, and maximizing building boundary area; Space - satisfying interior connectivity and adjacency, minimizing overlapping, satisfying orientation, minimizing overflow, maximizing compactness, satisfying area and size dimensions, and satisfying absolute position; and Openings - satisfying absolute position, satisfying orientation, minimizing overlapping, and satisfying minimum opening width and window-to-floor ratio. These performance indicators are aggregated in the objective function assessing the quality of solutions computed using the hybrid evolutionary strategy. The architectural elements are subject to geometric transformations, aiming to improve the performance indicators. The geometric transformation actions are to move, rotate, stretch, mirror, slice, and swap. These actions may be applied to single objects, such as openings or spaces, and to cluster of objects, such as all spaces in a floor level that share the same wall. The behavior of these transformations is adaptive, as they are more likely to be invoked according to their past success, i.e. improvement of the objective function, by selecting objects that are in nonconformity to the user's geometric and topologic specifications thus seeking to minimize the overall deviation, and by calculating the magnitude of the transformation (e.g. the distance and direction of an object translation). Figure 1 presents an example of a four-storey building model sided by two adjacent buildings. In each floor level, two apartments were set as building design specifications to be allocated within a building boundary.

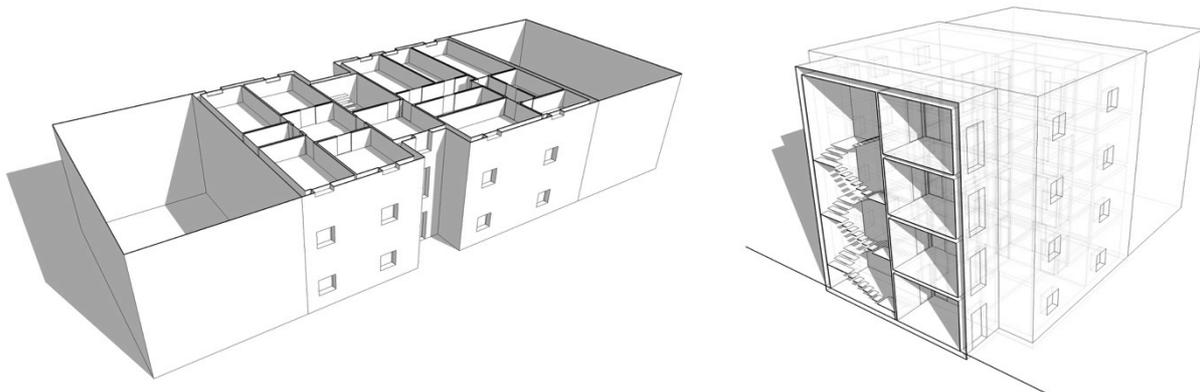

Figure 1: A four-storey building, sided by two adjacent buildings, generated by GerAPlanO prototype tool (model before thermal optimization). The design specifications have two apartments per floor level (one two-bedroom and one three-bedroom) served by a single stair. The 2D floor plans and the 3D model can be exported in a compatible DXF file and be opened in any CAD or BIM software. The model is cut through the second floor (left) and a section cut through the central stair and each level entrance is illustrated (right).

E. Rodrigues, A.R. Amaral, A.R. Gaspar, Á. Gomes, M.C.G. da Silva, and C.H. Antunes

## 2.2. Thermal assessment and optimization

The generated building designs can be evaluated using thermal performance criteria [9]. This is carried out using a coupled dynamic simulation engine that returns the operative temperature in each space with specifications of the number of occupants, their type of activity, and periods of use. The gains or losses due to infiltration and natural ventilation are also included in the assessment, as well as the internal gains due to the use of equipment and artificial lighting. The constructive system and location (and corresponding weather data) are specified by the user according to default data from a database or specifically set by him. The building elements (walls of spaces, ceilings, windows, doors, and pavements) may have their constructive layer materials set for a specific room without conflicting with the overall building construction solution.

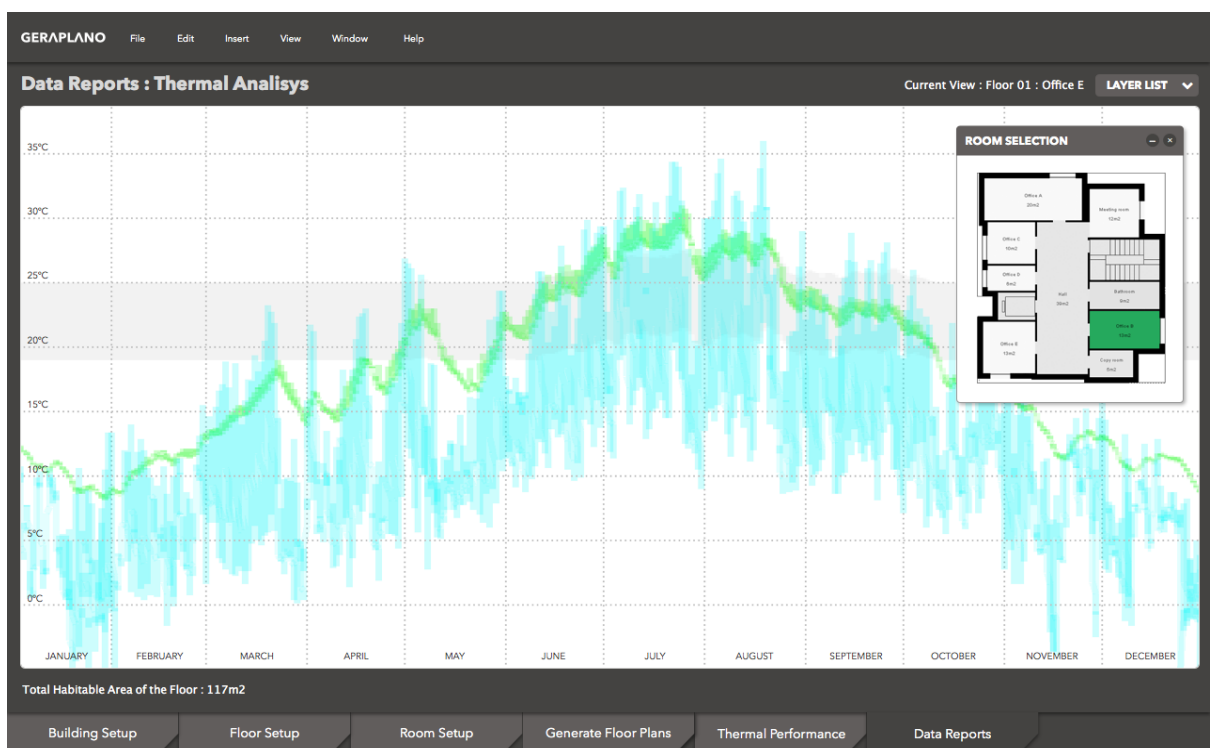

Figure 2: GerAPlanO layout window (beta version) considering the thermal comfort of a building space. The blue line represents outside air temperature and the green line displays the operative indoor temperature along the twelve months of the year. The grey area represents the human thermal comfort limits for the non-adaptive thermal comfort model.

The thermal performance optimization minimizes a weighted sum of cooling and heating degree-hours of thermal discomfort in all spaces. A sequential variable optimization procedure is used to improve the building performance according to the user design strategy [10]. The process consists in changing the geometry or/and adding shading mechanisms to the building design. The tool may change the building orientation, the position, size, and orientation of openings, walls position, reflect the floor plan, and determine the size of

44

overhangs and fins. A learning algorithm (e.g., based on neural networks) is being studied as an alternative estimation mechanism to dynamic simulation, hence shortening the runtime of the optimization process. Dynamic simulation will be used to obtain detailed information of the building performance and online training of the learning algorithm. Figure 2 depicts GerAPlanO window layout (beta version) considering the thermal behavior of a space after thermal comfort evaluation for naturally ventilated spaces according to the European Standard 15251:2007 (ANSI/ASHRAE Standard 55-2004 and non-adaptive thermal comfort limits are also implemented).

### 2.3. A web-based server system

The GerAPlanO prototype tool is being developed as a web-based server system where the user sets the overall project parameters (location, building boundary, maximum areas, and other building design specifications), generate alternative solutions, evaluate their thermal performance, and explore the improvement potential of each solution through the client browser. The user can download the 2D floor plan drawings or the 3D building model (see Figure 1) in compatible file types to be opened in conventional CAD and BIM programs, thus allowing the practitioner to further develop the building design. A web-based server system has the advantage of using the performance information, which is obtained every time that a user runs the dynamic simulation for a detailed building performance report, to train the learning algorithm and thus improving its estimation accuracy. Figure 3 illustrates the software layout to be presented.

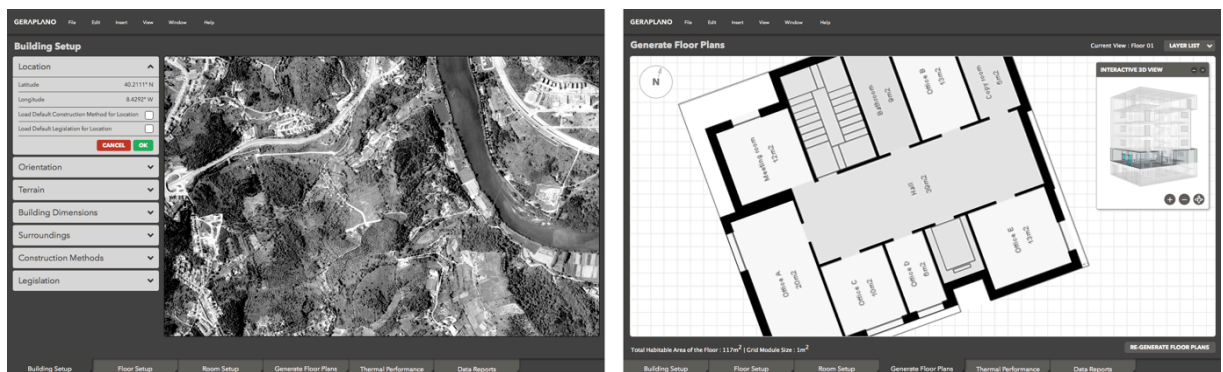

Figure 3: Examples of the GerAPlanO web-based layout (beta version). The building project may be geo-referenced and the building boundary drawn (left). Once the building solutions have been generated, the user may view the resulting floor plans and the corresponding 3D model (right).

### 3. CONCLUSION

The use of building performance simulation software can significantly contribute to the improvement of buildings' energy efficiency and thermal comfort. Different tools have been developed for this purpose; however, these tools generally increase the effort of an already long and difficult methodology of architectural design due to their computation burden. To overcome the difficulties in using dynamic simulation during building design process, the research underway in the GerAPlanO project combines, in an automatic process, the generation, evaluation and optimization of alternative design solutions in the space planning phase, in which building floor plans are drawn by architects, thus offering them a valuable interactive decision support tool. Experiments currently underway address real case studies with different levels of complexity, design programs, requirements and constraints, and will provide practitioners feedback related to the tool usability and utility, still offering researchers the opportunity to improve any shortcomings before the end of the project.

## ACKNOWLEDGEMENTS

This work has been developed under the *Energy for Sustainability Initiative* of the University of Coimbra (UC) and supported by the project *Automatic Generation of Architectural Floor Plans with Energy Optimization* - GerAPlanO - QREN 38922 Project (CENTRO-07-0402-FEDER-038922).

## REFERENCES

[1] IEA, Tracking Clean Energy Progress 2014, Paris, France, 2014. http://www.iea.org/etp/tracking/.
[2] J. Whyte, Towards a new craft of architecture, Build. Res. Inf. (2014) 1–3. doi:10.1080/09613218.2015.962240.
[3] J. Kanters, M. Horvat, M.-C. Dubois, Tools and methods used by architects for solar design, Energy Build. 68 (2014) 721–731. doi:10.1016/j.enbuild.2012.05.031.
[4] GerAPlanO Project, (2013). http://www.uc.pt/en/efs/research/geraplano.
[5] E. Rodrigues, Automated Floor Plan Design: Generation, Simulation, and Optimization, University of Coimbra, 2014. http://hdl.handle.net/10316/25438 (accessed July 24, 2014).
[6] E. Rodrigues, A.R. Gaspar, Á. Gomes, An evolutionary strategy enhanced with a local search technique for the space allocation problem in architecture, Part 1: Methodology, Comput. Des. 45 (2013) 887–897. doi:10.1016/j.cad.2013.01.001.
[7] E. Rodrigues, A.R. Gaspar, Á. Gomes, An evolutionary strategy enhanced with a local search technique for the space allocation problem in architecture, Part 2: Validation and performance tests, Comput. Des. 45 (2013) 898–910. doi:10.1016/j.cad.2013.01.003.
[8] E. Rodrigues, A.R. Gaspar, Á. Gomes, An approach to the multi-level space allocation problem in architecture using a hybrid evolutionary technique, Autom. Constr. 35 (2013) 482–498. doi:10.1016/j.autcon.2013.06.005.
[9] E. Rodrigues, A.R. Gaspar, Á. Gomes, Automated approach for design generation and thermal assessment of alternative floor plans, Energy Build. 81 (2014) 170–181. doi:10.1016/j.enbuild.2014.06.016.
[10] E. Rodrigues, A.R. Gaspar, Á. Gomes, Improving thermal performance of automatically generated floor plans using a geometric variable sequential optimization procedure, Appl. Energy. 132 (2014) 200–215. doi:10.1016/j.apenergy.2014.06.068.